\input harvmac
\input epsf

\overfullrule=0pt
\abovedisplayskip=12pt plus 3pt minus 3pt
\belowdisplayskip=12pt plus 3pt minus 3pt
\sequentialequations

\lref\dienes{K.R. Dienes, {\it ``Understanding gauge coupling
unification in string theory: A review''},
Nucl. Phys. Proc. Suppl. {\bf 52A} (1997), 276.}
\lref\dienrev{K.R. Dienes,
{\it ``String theory and the path to unification: 
A review of recent developments''},
hep-th/9602045,
Phys. Rep. {\bf 287} (1997), 447.}
\lref\gsw{M.B. Green, J.H. Schwarz and E. Witten, {\it ``Superstring
Theory''} Vol 1 and 2, Cambridge University Press (1987).}
\lref\witcy{E. Witten, 
{\it ``Strong coupling expansion of Calabi-Yau compactification''},
hep-th/9602070,
Nucl. Phys. {\bf B471} (1996) 135.}
\lref\sendual{A. Sen, {\it ``Dyon-monopole bound states, selfdual
harmonic forms on the multi-monopole moduli space, and SL(2,Z)
invariance in string theory''},
hep-th/9402032,
Phys. Lett. {\bf B329} (1994) 217.}
\lref\vafwit{C. Vafa and E. Witten,
{\it ``A strong coupling test of S duality''},
hep-th/9408074,
Nucl. Phys. {\bf B431} (1994), 3.}
\lref\seiwit{N. Seiberg and E. Witten,
{\it ``Monopoles, duality and chiral symmetry breaking in N=2
supersymmetric QCD''},
hep-th/9408099,
Nucl. Phys. {\bf B431} (1994), 484;
{\it ``Electric-magnetic duality, monopole condensation, and
confinement in N=2 supersymmetric Yang-Mills theory''},
hep-th/9407087,
Nucl. Phys. {\bf B426} (1994), 19.}
\lref\senf{A. Sen, {\it ``F-theory and orientifolds''}, 
hep-th/9605150, Nucl.  Phys. {\bf B475} (1996), 562.}
\lref\bds{T. Banks, M. Douglas and N. Seiberg, 
{\it ``Probing F theory with branes''}, hep-th/9605199,
Phys. Lett. {\bf B387} (1996), 278.}
\lref\townm{P.K. Townsend, {\it ``The eleven-dimensional supermembrane 
revisited''}, hep-th/9501068, Phys. Lett. {\bf B350} (1995), 184.}
\lref\witvar{E. Witten, {\it ``String theory dynamics in various
dimensions''}, hep-th/9503124, Nucl. Phys. {\bf B443} (1995), 85.}
\lref\schwarzm{J.H. Schwarz, {\it ``The power of M-theory''},
hep-th/9510086, Phys. Lett. {\bf B367} (1996), 97.}
\lref\aspin{P. Aspinwall, 
{\it ``Some relationships between dualities in string theory''},
hep-th/9508154,
Nucl. Phys. Proc. Suppl. {\bf 46} (1996) 30.}
\lref\horwit{P. Horava and E. Witten, {\it ``Heterotic and type I 
string dynamics from eleven-dimensions''}, hep-th/9510209,
Nucl. Phys. {\bf B460} (1996), 506.}
\lref\dmm{K. Dasgupta and S. Mukhi, {\it ``Orbifolds of M-theory''},
hep-th/9512196, Nucl. Phys. {\bf B465} (1996), 399.}
\lref\witfive{ E. Witten, {\it ``Five branes and M-theory on an 
orbifold''}, hep-th/9512219, Nucl. Phys. {\bf B463} (1996), 383.}
\lref\senm{A. Sen, {\it ``M theory on $(K3 \times S^1)/Z_2$''},
hep-th/9602010,
Phys. Rev. {\bf D53} (1996), 6725.}
\lref\gopm{R. Gopakumar and S. Mukhi, 
{\it ``Orbifold and orientifold compactifications of F-theory 
and M-theory to six dimensions and four dimensions''},
hep-th/9607057,
Nucl. Phys. {\bf B479} (1996) 260.}
\lref\gimjohn{E. Gimon and C. Johnson, {\it ``Multiple Realizations
of N=1 Vacua in Six-Dimensions''}, hep-th/9606176,
Nucl. Phys. {\bf B479} (1996), 285.}
\lref\kaplun{E. Caceres, V.S. Kaplunovsky and I.M. Mandelberg, {\it 
``Large-volume string compactifications, revisited''}, hep-th/9606036,
Nucl. Phys. {\bf B493} (1997), 73.}
\lref\partlist{J. Ellis, A.E. Faraggi and D.V. Nanopoulos,
{\it ``M theory model building and proton stability''},
hep-th/9709049\semi
E. Dudas,
{\it ``Supersymmetry breaking in M theory and quantization rules''},
hep-th/9709043\semi
I. Antoniadis and M. Quiros,
{\it ``Supersymmetry breaking in M theory''},
hep-th/9709023;
{\it ``Supersymmetry breaking in M theory and gaugino condensation''},
hep-th/9705037;
{\it ``Large radii and string unification''},
hep-th/9609209,
Phys. Lett. {\bf B392} (1997), 61\semi
V.~Kaplunovsky and J.~Louis,
{\it ``Phenomenological aspects of F theory''},
hep-th/9708049\semi
A. Brignole, L.E. Ibanez and C. Munoz,
{\it ``Soft supersymmetry breaking terms from supergravity and superstring
models''},
hep-ph/9707209\semi
K. Choi,
{\it ``Axions and the strong CP problem in M theory''},
hep-th/9706171\semi
G. Aldazabal, A. Font, L.E. Ibanez, A.M. Uranga and G. Violero,
{\it ``Non-perturbative heterotic D = 6, D = 4, N=1 orbifold vacua''},
hep-th/9706158\semi
E. Dudas and Christophe Grojean,
{\it ``Four-dimensional M theory and supersymmetry breaking''},
hep-th/9704177\semi
T. Li, J.L. Lopez and D.V. Nanopoulos,
{``Compactifications of M theory and their phenomenological 
consequences''},
hep-ph/9704247,
Phys. Rev. {\bf D56} (1997), 2602;
{\it ``M theory inspired no scale supergravity''},
hep-ph/9702237\semi
E. Dudas and J. Mourad,
{\it ``On the strongly coupled heterotic string''},
hep-th/9701048,
Phys. Lett. {\bf B400} (1997) 71\semi
P. Horava,
{\it ``Gluino condensation in strongly coupled heterotic string theory''}
hep-th/9608019,
Phys. Rev. {\bf D54} (1996), 7561\semi
E. Kiritsis, C. Kounnas, P.M. Petropoulos and J. Rizos, 
{\it ``Solving the decompactification problem in string theory''},
hep-th/9606087,
Phys. Lett. {\bf B385} (1996), 87.}
\lref\vafaf{C. Vafa, {\it ``Evidence for F-theory''}, hep-th/9602022,
Nucl. Phys. {\bf B469} (1996), 403.}

{\nopagenumbers
\Title{\vtop{\hbox{hep-ph/9710470}
\hbox{TIFR/TH/97-55}}}
{\vtop{
\centerline{Recent Developments in String Theory:}
\medskip
\centerline{A Brief Review for Particle Physicists}}}
\footnote{}{Based on an invited talk given at the XII DAE Symnposium
on High Energy Physics, Guwahati, December 1996}

\centerline{Sunil Mukhi\foot{E-mail: mukhi@theory.tifr.res.in}}
\vskip 4pt
\centerline{\it Tata Institute of Fundamental Research,}
\centerline{\it Homi Bhabha Rd, Bombay 400 005, 
India}
\ \medskip
\centerline{ABSTRACT}

At the present time, string theory (and its generalizations) remain
relatively abstruse subjects to the particle phenomenologist and
experimentalist. Yet, striking developments of the last two years
offer hope that a fundamental non-perturbative formulation of this
theory will be found, and that this formulation will permit us to make
contact with supersymmetric standard-model physics. This article is
based on a talk which attempted to convey the essence of these recent
developments in string theory, in a non-technical manner, to an
audience of particle theorists, phenomenologists and experimentalists.

\ \vfill 
\leftline{October 1997}
\eject}
\ftno=0

\newsec{Introduction: Historical Perspective}

The theory of relativistic quantized strings became a compelling
candidate for the unification of elementary interactions, after a
number of remarkable properties were uncovered. This process took over
a decade, at the end of which (around 1983) one could make the
following list of properties\foot{For all references to these 
results of the pre-duality era, see Ref.\refs{\gsw}}:
\medskip

\item{*} At low energies, string theory is described by a 
supersymmetric quantum field theory containing gravitons, gauge
particles and fermions in a unified framework.

\item{*} At high energies, it differs from ordinary field theory, and 
is ultraviolet finite {\it despite} the presence of gravitons.

\item{*} There is no coupling constant. Instead, there is a massless 
scalar field, the ``dilaton'', whose (perturbatively undetermined) VEV
acts as the gauge and gravitational coupling.

\item{*} Supersymmetry seems to be {\it required} for consistency.

\item{*} The total number of spacetime dimensions is 10, although as
in any theory containing gravity, each of these dimensions could be
compact or noncompact.  A {\it choice of vacuum configuration} is
required for a physical interpretation as spacetime, and this could
have any number $D \le 10$ of noncompact spacetime dimensions.
\medskip

Remarkably, this list of properties failed to cause large-scale
enthusiasm in the theoretical particle physics community. But
unprecedented excitement arose in 1984 when one more impressive and
rather unexpected property was added to this list. The superstrings
with $N=1$ supersymmetry in 10 dimensions, which are the ones
possessing Yang-Mills gauge fields, are chiral and hence potentially
anomalous at the quantum level. It was shown that they are in fact
free of gauge and gravitational anomalies for just two choices of the
gauge group: $SO(32)$ and $E_8\times E_8$.

This sparked off a wave of interest in string theory, leading to the
discovery of a whole new set of properties:
\medskip

\item{*} The superstring with $E_8\times E_8$ gauge symmetry, the
``heterotic string'', admits consistent vacuum solutions of the form of a
compact 6-manifold, times flat 4D Minkowski spacetime. A natural class
of such models give $N=1$ supersymmetry in 4D. This is only a quarter
as much SUSY as $N=1$ in 10D, as the 6-manifold background -- a
``Calabi-Yau'' space -- breaks $3/4$ of the original supersymmetries.

\item{*} This class of models easily admits chiral fermion 
representations in 4d. A simple pattern of symmetry-breaking is 
\eqn\symmbreak{
E_8\times E_8 \rightarrow E_8 \times E_6}
with fermions in the $\underline{27}$ of $E_6$. Alternatively $E_6$
could be replaced by $SO(10)$ with fermions in the $\underline{16}$. 

\item{*} The net number of chiral fermion generations is related to the
topology of the Calabi-Yau 6-manifold by, for example,
\eqn\netno{
n_{27} - n_{\overline{27}} = \half |\chi| }
where $\chi$ is the Euler characteristic of the manifold.

\item{*} In this scenario, one entire $E_8$ is naturally unbroken, but
communicates only gravitationally with the standard gauge sector. Thus
it could plausibly be interpreted as a ``hidden sector'' consisting of
dark matter.
\medskip

The above picture is grossly oversimplified in many
ways. Compactifications far more general than the type described above
are equally possible, but require more complicated techniques for
their analysis. More important, the above results did not provide much
of a clue about what should be done to give the 4-dimensional theory a
truly realistic spectrum of particles: this would require
supersymmetry to be broken, and both masses and mass splittings to be
generated. The dilaton would need to be stabilized by the generation
of a potential.

In addition, there was one problematic prediction in the heterotic
string. This can be shown in various ways, though we will use a
simple, classical scaling argument due to Witten\refs{\witcy}. The
Planck scale in a $D$-dimensional theory of gravity is defined (upto
constants of order 1) by writing the action as
\eqn\plancksc{
S_D[g] = (M^{pl}_D)^{D-2}\int d^D x\, \sqrt{g} R }
where the power of $M^{pl}_D$ makes the action dimensionless, given
that the action is built from dimensionless fields, 2 derivatives and
$D$ integrations. In 10-dimensional string theory the low-energy
effective action action is also commonly written
\eqn\tendgrav{
S_{10}[g] = (M_s)^8 \int d^{10}x\, e^{-2\phi} \sqrt{g} R }
where $M_s$ is by definition the string scale, and $\phi$ is the
dilaton, related to the string coupling by $g_s = e^\phi$. From this
it is evident that the 10d Planck mass in string theory satisfies
$(M^{pl}_{10})^8 = (g_s)^{-2} (M_s)^8$.

Now if we dimensionally reduce the above actions on a 6-dimensional
manifold of volume $V$ (assumed to be isotropic), the volume factors
out and the effective 4d action can be written:
\eqn\fourdgrav{
\eqalign{
S_4[g] &= V (M^{pl}_{10})^{8}\int d^4 x\, \sqrt{g} R\cr
&= (M^{(pl)}_{4})^2 \int d^4 x\, \sqrt{g} R }}
From the above equations, the 4-dimensional Planck mass is related to
the string parameters $M_s,g_s,V$ by
\eqn\fourdpl{
M^{(pl)}_{4} \sim \sqrt{V}\, {(M_s)^4\over g_s} }

A similar calculation can now be carried out for the gauge
interactions arising from string theory. Here, we require that the
gauge fields have canonical dimension 1 in 10 dimensions, so that upon
dimensional reduction to 4 dimensions they will automatically have the
canonical dimensions appropriate to 4d gauge theory. Thus the string
low-energy action in 10d has a term\foot{The dilaton-dependence in
this term is specific to the heterotic string.}
\eqn\tendgauge{
S_{10}[A] = (M_s)^6 \int d^{10}x\, e^{-2\phi} \sqrt{g}\, 
\tr F_{\mu\nu} F^{\mu\nu} }
After compactification, the 4d action will now be interpreted as the
GUT action for the gauge theory, so it is written
\eqn\fourdgauge{
\eqalign{
S_4[A] &= V {(M_s)^6 \over (g_s)^2} \int d^4 x\, \sqrt{g}\, 
\tr F_{\mu\nu} F^{\mu\nu}\cr
&= {1\over (\alpha_{\rm GUT})^2} \int d^4 x\, \sqrt{g}\, 
\tr F_{\mu\nu} F^{\mu\nu}}}
from which we get our second equation, relating the 4d GUT coupling
constant to the string parameters:
\eqn\fourdgut{
{1\over \sqrt{\alpha_{\rm GUT}}} \sim \sqrt{V}\, {(M_s)^3\over g_s} }

To get a prediction about the real world, we must eliminate the string
scale $M_s$ between Eqs.\fourdpl\ and \fourdgut, and set the volume
$V$ of the internal space to be given by the GUT mass scale: $V\sim
(M_{\rm GUT})^{-6}$. Then we find:
\eqn\prob{
M^{(pl)}_4 \sim (g_s)^{1/3} {M_{\rm GUT}\over \alpha_{\rm GUT}^{2/3}} }
For weakly coupled string theory, the above relation implies
\eqn\inequal{
M^{(pl)}_4 \ll {M_{\rm GUT}\over \alpha_{\rm GUT}^{2/3}} }
which is unacceptably small. This is the gauge coupling unification
problem in string theory.

While many ways have been proposed to overcome this problem (see for
example Ref.\refs{\dienes} and references therein), it is clear that
the problem goes away if we are not committed to weakly coupled string
theory. In other words, if we know something about the strongly
coupled, nonperturbative regime in string theory, then Eq.\prob\ above
can simply be used as a phenomenological input to determine what is
the (large) value of the string coupling relevant to the real world.

Recent developments in string theory have made this seem less of an
impossibility, and we will return to this point.

\newsec{Dualities in String Theory}

Two kinds of duality have played a crucial role in enlarging our
understanding of string theory. I will briefly review how they
operate. 

The first duality that we encounter is called ``target-space duality''
or ``T-duality''. This is an essentially stringy phenomenon, but can
be understood in a non-technical way from a simple picture.

Consider a spacetime containing one direction compactified on a circle
of radius $R$. A point particle propagating on the circle has
quantized momenta $p\sim {n\over R}$ , as we know from elementary
quantum mechanics. A string propagating on the same circle can behave
like a point particle located at its centre of mass (just by becoming
a very small string), so string theory too has quantized momentum
modes on a circle, with energies $E\sim {n\over R}$.

But the string can do something new: it can ``wrap'' itself on the
circle an integer number of times, creating a ``winding mode''. These
modes have an energy proportional to the radius, simply because the
energy in an extended string is equal to its tension (a constant of
string theory) times its length. Thus for such modes $E\sim m (M_s)^2
R$ where $m$ is some integer, and the string scale has been inserted
for dimensional reasons. The two kinds of modes are depicted in this
figure: 
\bigskip

\centerline{\epsfbox{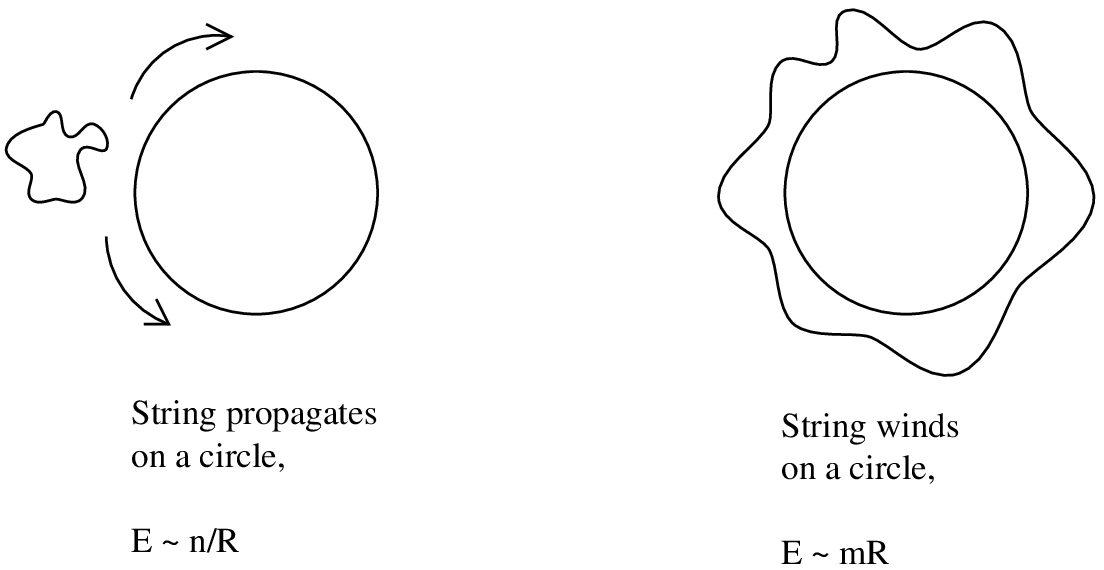}}
\bigskip

Now, the replacement
\eqn\tdual{
M_s R \rightarrow {1\over M_s R} }
interchanges the spectra of the string momentum and winding
modes. Since this operation can be carried out at weak coupling, one
can ask if it is a symmetry of perturbative string theory. Indeed, it
can be shown to be so (we will not do it here).

Thus, in string theory, short (sub-Planckian) distances are equivalent
to long distances. As one consequence, a string theory compactified on
a very small circle will be equivalent to one on a very large circle,
hence the number of (approximately) noncompact spacetime dimensions in
a given situation is not well-defined. The winding modes on a small
circle behave like momentum modes on a large one. Our physical
interpretation, based on particles, would probably be in terms of the
larger number of dimensions.

Another duality, for which the arguments are by no means so rigorous,
is ``strong-weak duality'' or ``S-duality''. This is the statement
that, in some cases, weakly coupled string theory with string coupling
$g_s$ is equivalent to strongly coupled string theory with coupling
${1\over g_s}$. As with T-duality, different types of modes get
interchanged by the S-duality operation. In some situations (typically
in compactifications to four spacetime dimensions) it exchanges
electrically charged with magnetically charged excitations of the
theory.

More generally, strong-weak duality exchanges ``fundamental'' with
``solitonic'' modes in string theory. The former are perturbative
while the latter are visible in the spectrum as non-trivial solutions
of the classical equations of motion. (More details about this were
given in the lecture of A. Dabholkar at this Symposium.)

Let us investigate S-duality a little more explicitly, in the most
basic situation where it appears. This is the so-called type IIB
superstring in 10 dimensions, whose low-energy Lagrangian is type IIB
(i.e. chiral, $N=2$) supergravity. The spectrum of massless fields in
this supergravity is dictated purely by classical considerations. The
only irreducible supermultiplet is that of supergravity, with bosonic
fields as follows:
\eqn\iibspect{
\eqalign{
g_{\mu\nu}&:\qquad {\rm metric}\cr
B_{\mu\nu}&:\qquad {\rm 2-form~field}\cr
{\tilde B}_{\mu\nu}&: \qquad {\rm another~2-form~field}\cr
\phi&: \qquad {\rm scalar~(dilaton)}\cr
{\tilde\phi}&: \qquad {\rm another~scalar~(axion)}\cr
{\tilde D}_{\mu\nu\lambda\rho}&: \qquad {\rm 4-form~field}}}
Here, by ``$p$-form field'' we mean a bosonic field transforming as a
totally antisymmetric $p$-th rank tensor. This multiplet contains
fermions too, but we will not need to list them.

It is convenient to define the complex scalar field $\tau \equiv
{\tilde\phi} + i e^{-\phi}$. The low-energy Lagrangian, in the
approximation of slowly-varying fields (no more than two derivatives),
is dictated by supersymmetry and turns out to be invariant under
\eqn\sltwor{
\tau \rightarrow {a\tau + b\over c\tau + d}\qquad
\pmatrix{B\cr {\tilde B}} \rightarrow \pmatrix{a &c\cr b &d}
\pmatrix{B\cr {\tilde B}} }
where $a,b,c,d$ are real numbers satisfying $ad-bc=1$. These
transformations form the group $SL(2,R)$. 

Although this is a symmetry of the low-energy effective Lagrangian, it
certainly cannot be an exact symmetry of string theory. The reason is
that string theory has states carrying {\it charge} under the 2-form
fields $B, {\tilde B}$. Indeed, such states are strings: in general,
an object extended in $p-1$ spatial directions carries charge under
$p$-forms. The familiar case is $p=1$ for which we get particles
charged under gauge fields. The type IIB string itself carries unit
charge under $B_{\mu\nu}$, while there is a certain solitonic string
carrying unit charge under ${\tilde B}_{\mu\nu}$. $SL(2,R)$ would not
only mix up these charges but in general make them fractional (even
irrational) which is forbidden by charge quantization.

The largest subgroup of $SL(2,R)$ which can conceivably be an exact
symmetry of the IIB string is $SL(2,Z)$, consisting of matrices
$\pmatrix{a &b\cr c&d}$ with $a,b,c,d$ integral and $ad-bc=1$. This
preserves integrality of charges, but, from Eq.\sltwor, rotates the
$B$ and ${\tilde B}$ fields into each other, hence also the
fundamental into the solitonic string -- as promised.

A large and very tightly constrained set of results leads us to
believe that $SL(2,Z)$ is indeed an exact symmetry of the type IIB
string. In particular, the choice of transformation $\pmatrix{0 &-1\cr
1 &0}$ corresponds to the transformation
\eqn\stransf{
\tau \rightarrow -{1\over \tau} }
which, if ${\tilde\phi}=0$, corresponds to
\eqn\swduality{
e^\phi \rightarrow e^{-\phi} }
or in other words, $g_s \rightarrow g_s^{-1}$. Thus the S-duality
group $SL(2,Z)$ contains strong-weak duality as a $Z_2$ subgroup. The
non-Abelian nature of the full group, however, makes it still more
interesting.

Although not strictly related to the theme of this talk, this stringy
duality has some amazing implications for ordinary (supersymmetric)
quantum field theory without gravity. In recent years, it has been
argued with considerable difficulty and
ingenuity\refs{\sendual,\vafwit} that $N=4$ supersymmetric Yang-Mills
(SYM) theory in $3+1$ dimensions has exact electric-magnetic $SL(2,Z)$
duality symmetry. It has also been argued\refs{\seiwit} that $N=2$ SYM
field theory can be solved nonperturbatively, and that this solution
makes use of $SL(2,Z)$ duality transformations in an essential way.

Both these facts, superficially unrelated to each other and to
10-dimensional string theory, are actually consequences of the
$SL(2,Z)$ duality of the type IIB string that we have discussed
above. Type IIB string theory contains various extended objects
including ``3-branes'' and ``7-branes'' (similar to the membranes or
domain walls studied in cosmological theories, but one and five
dimensions higher respectively). It turns out that the dynamics of an
isolated 3-brane is governed by an $N=4$ SYM field theory, which
inherits the $SL(2,Z)$ duality from the string theory in which it is
embedded. Similarly, the dynamics of a 3-brane parallel to, and near,
a 7-brane, is governed by an $N=2$ SYM field theory, whose
nonperturbative solution, including the duality transformations, is
inherited from the type IIB string\refs{\senf,\bds}.

One lesson from this is that nontrivial properties of flat-spacetime
field theories become more transparent when these theories are viewed
as sectors of string theory. Even if string theory is not the right
unified theory of nature, it could well be an essential new way to
understand quantum field theories!

\newsec{Beyond String Theory: ``M-Theory''}

We have seen that the type IIB string has a strong-weak-coupling
S-duality. This means that its dynamics at strong coupling is
understood, and can be extracted by mapping onto a weakly coupled dual
theory. However, an analogous result cannot hold for the type IIA
string, whose low-energy limit is non-chiral $N=2$ supergravity. The
low energy effective action in 10 spacetime dimensions does not admit
any symmetry which includes interchange of strong and weak
coupling. So a new insight is needed.

In the old days, extended supergravity theories were often constructed
by compactifying higher dimensional theories on a circle (the
``Kaluza-Klein'' procedure). For certain values of $D$, spinors of the
Lorentz algebra split into two when we go down from $D$ to $D-1$
dimensions. In these cases, compactification of a supersymmetric
theory doubles the number of spinor supercharges. Moreover, if $D$ is
odd, then the parent theory is obviously non-chiral, and reduction
leaves us with a non-chiral theory in even dimensions.

In 11 spacetime dimensions, there is a unique classical supergravity
theory, which has $N=1$ supersymmetry. Upon Kaluza-Klein reduction on
a circle, it gives precisely the type IIA supergravity in $D=10$. The
11-dimensional theory possesses no scalar particle analogous to the
dilaton of string theory, hence it has no dimensionless coupling
constant. But after circle compactification, one component of the
metric becomes such a scalar. By comparing classical actions in 10D
and 11D, it is easy to obtain the following relations between
11-dimensional quantities (the Planck mass $M^{(pl)}_{11}$ and the
compactification radius $R_{11}$) and 10-dimensional quantities (the
string scale and the string coupling):
\eqn\relns{
\eqalign{
M^{(pl)}_{11} &= {M_s\over (g_s)^{1\over 3}}\cr
R_{11} &= {g_s\over M_s}\cr}}
Now if we choose units for which $M^{(pl)}_{11}=1$ then we find
\eqn\eqnforr{
R_{11} = (g_s)^{2\over 3} = e^{{2\over3}\phi} }

Classically, this seems to say that type IIA supergravity in $D=10$
becomes 11-dimensional at strong coupling. But there is considerable
evidence for a much stronger conjecture: that the strong-coupling
limit of {\it quantized} type IIA {\it string} theory, is some ``new''
11-dimensional quantum theory, having 11-dimensional supergravity as
its low-energy limit. This new theory is called
``M-theory''\refs{\townm,\witvar}.

Thus, M-theory is more than just supergravity in $D=11$. For example,
it contains membranes in its spectrum. Their presence explains the
relation to string theory: the type IIA string is believed to be just
the membrane of M-theory, wrapped round the circle of radius
$R_{11}$. This looks like a string for small values of $R_{11}$, which
by Eq.\eqnforr\ above, is just the limit of weakly coupled string
theory where we indeed expect to see the fundamental string. The
situation looks as follows:

\bigskip

\centerline{\epsfbox{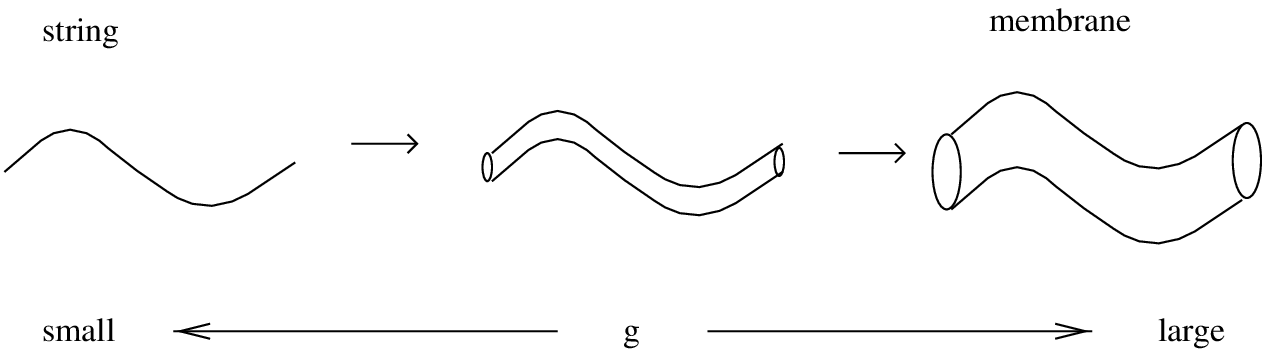}}
\bigskip

Some people believe that M-theory is a theory of fundamental
membranes, rather than strings, but this has yet to be convincingly
demonstrated. It has, however, very convincingly been argued that
M-theory is a sensible quantum theory containing various extended
objects. The very existence of this theory implies many important
properties of string theory.  For example, the type IIA string is
known to contain other extended objects besides strings, the
``$p$-branes'', and these can be argued to arise from
M-theory\refs{\schwarzm}.

Even more impressive, one can relate M-theory to the type IIB string,
though the latter is a chiral theory. The relation is that M-theory
compactified to 9 dimensions on a 2-torus, in the limit that the torus
shrinks, is the 10-dimensional type IIB string!  To argue this
equivalence\refs{\schwarzm,\aspin}, we make a T-duality on the way,
which interchanges the type IIA and type IIB strings. The apparent
paradox inherent in equating a 9-dimensional with a 10-dimensional
theory is resolved by the observation, made earlier, that T-duality in
string theory permits precisely this.

As a consequence, we can {\it predict} the nonperturbative $SL(2,Z)$
S-duality of the type IIB string. It is just the geometrical $SL(2,Z)$
symmetry group of the 2-torus on which M-theory is compactified!
Strong-weak coupling interchange is just exchange of the two cycles of
the torus. Thus M-theory plays a powerful role in geometrizing the
dualities of string theory.

\newsec{M-theory and the Heterotic String}

So far we have spoken of M-theory as an 11-dimensional theory
underlying the type IIA and IIB strings. But these are not the string
theories having the most promising connection with particle physics.
As we discussed in the beginning, it is the $E_8\times E_8$ heterotic
string which has the most plausible relationship to grand
unification. I will now argue that the $E_8\times E_8$ heterotic
string can also be derived from M-theory\refs{\horwit}.

Although M-theory is less well-understood than string theory, it has
some ``stringy'' properties. Among these is the fact that it can be
compactified not only on smooth manifolds (which elementary particle
theories require) but also on singular ones called orbifolds. An
orbifold is an almost-everywhere smooth manifold with some specific
types of singularities. It can be described as the quotient $M/\Gamma$
where $M$ is a smooth manifold (typically a torus) and $\Gamma$ is a
discrete group of symmetries of $M$. The quotient space has
singularities wherever the quotienting group has fixed points on the
manifold.

The spectrum of strings on an orbifold $M/\Gamma$ is related to that
on $M$. First of all, clearly we must keep string configurations on
$M$ which are invariant under $\Gamma$. But additionally, we should
include string configurations which would not have been well-defined
on $M$, but are well-defined after quotienting with $\Gamma$. The
latter are called ``twisted sector states''. 

The simplest orbifold possible comes from the choice $M=S^1$, a
circle, and $\Gamma = Z_2$. Let the circle coordinate $x$ take the
range $0\le x \le 2\pi R$, and let the $Z_2$ action send $x$ to
$-x$. The quotient space is the interval $0\le x\le\pi R$ with two
end-points. Compactifying M-theory to 10 dimensions on this simple
orbifold leads to a remarkable surprise.

It can be shown that the $Z_2$ action projects out half the gravitinos
that M-theory would have had if compactified just on a circle to
10D. Thus it has only $N=1$ supersymmetry rather than $N=2$. Indeed,
the $Z_2$ projection leaves us with the $N=1$ supergravity multiplet
as the ``untwisted'' sector. If there is an analogue of a twisted
sector, then this is restricted by supersymmetry, which admits only a
super-Yang-Mills (SYM) multiplet besides the supergravity
multiplet. So without any computation, we conclude that the twisted
sector, if it gives anything, contributes Yang-Mills fields and hence
gauge symmetry to the orbifold theory.

With our limited understanding of M-theory it would be hard to predict
the actual gauge group, but for one happy circumstance. The $N=1$
supergravity multiplet in 10D is chiral and has a gravitational
anomaly. Therefore, to get a consistent theory we must find a way of
cancelling the anomaly using the twisted sector states. 

Now, it has long been known that gravitational, gauge and mixed
anomalies can be cancelled in $N=1$ supergravity in 10D only if the
gauge group is one of $SO(32)$ or $E_8\times E_8$. But in our case the
anomalies are located at the two ends of the interval $0\le x\le \pi
R$ (in between, the theory is effectively 11-dimensional and cannot
have gravitational anomalies). So local cancellation of anomalies
requires that the gauge group appear in two factors, one associated to
each end of the interval. That uniquely fixes the gauge group, from
among the two choices above, to be $E_8\times E_8$.

This leads to the conjecture that M-theory compactified on $S^1/Z_2$
is actually the heterotic string with gauge group $E_8\times
E_8$. This immediately brings M-theory into prominence in attempts to
extract particle physics as a low-energy limit of string theory. When
the string is strongly coupled, an 11th dimension, in the form of an
interval with two end-points, opens up. 

More general orbifolds of M-theory also exist\refs{\dmm,\witfive,\senm,
\gopm,\gimjohn}, and should play an important role in providing more
general classes of M-theory compactification to 4 dimensions.

\newsec{An M-theory Application to Unification Physics}

With this result, the strong coupling behaviour of the heterotic
string is no longer a mystery. At strong coupling, we replace the
perturbative string description (which is no longer appropriate) by
the description as M-theory on 10D spacetime times an interval
$S^1/Z_2$. The length of the interval is proportional to a power of
the string coupling, exactly as in Eq.\eqnforr. 

The coupling unification problem that was described in Section 1 now
admits an obvious solution. We are no longer committed to the weakly
coupled heterotic string, so the undesirable bound Eq.\inequal\ need
not hold. Instead, one can use Eqs.\fourdgut,\prob\ and \relns\ to
{\it calculate} the radius of the 11th dimension. One
finds\refs{\witcy}:
\eqn\calcrad{
M^{(pl)}_4 \sim {R_{11}^{1\over2} M_{\rm GUT}^{3\over2} \over 
\alpha_{\rm GUT}^{3\over 4}}}
from which the radius of the eleventh dimension is
\eqn\solvrad{
R_{11} \sim {(M^{(pl)}_4)^2 \alpha_{\rm GUT}^{3\over2} \over M_{\rm
GUT}^3} }

A more careful calculation leads to an estimate of around 80 for the
numerical coefficient on the RHS of Eq.\calcrad, hence the RHS of
Eq.\solvrad\ actually has a prefactor as small as $1.5 \times
10^{-4}$. 

A detailed analysis of this and other fascinating results in the
literature is beyond the scope of the present article. It should be
noted that the present scenario is far from complete or convincing,
for a variety of reasons (see, for instance, the very detailed
discussion in Ref.\refs{\kaplun}). The purpose of exhibiting these
simple calculations here is to illustrate how problems that were
virtually impossible to address in perturbative string theory, have
been opened up for discussion with the advent of string dualities and
M-theory.

\newsec{Conclusions}

String theory remains the prime candidate for a unified theory of all
fundamental interactions. Recent developments have given us unexpected
control over some nonperturbative and strong-coupling effects in
string theory. Moreover, it appears that whatever we formerly knew
about string theory, plus much more, can be subsumed into a more
general framework called ``M-theory''.

A modest literature has grown up over the proposal that 4-dimensional
unification physics can be obtained from compactifications of
M-theory, either of the type $S^1/Z_2\times M_6$ discussed above, or
directly via more general orbifolds\refs{\gopm,\gimjohn}. The
interested reader may consult the review by Dienes\refs{\dienrev} and
the papers in Ref.\refs{\partlist} (this is only a partial list of
recent papers).

Due to a shortage of spacetime, an important and complementary
development called ``F-theory''\refs{\vafaf} has been left out of the
present discussion. F-theory offers another approach to study
nonperturbative string physics after compactification to 4
dimensions. After years of having no control over strong-coupling
behaviour in compactified string theories, we now have at least two
different approaches! It remains to be seen how long it takes to
assemble all this into genuine progress towards the goal of
unification.

\listrefs
\end